\documentclass[12pt]{article}
\usepackage{amsmath,epsfig,euscript,url}

\usepackage{bm}

\def\pslash{\rlap{\hspace{0.02cm}/}{p}}
\def\kslash{\rlap{\hspace{0.02cm}/}{k}}
\def\lslash{\rlap{\hspace{-0.02cm}/}{l}}

\begin{document}

\begin{titlepage}

\begin{flushright}
WSU-HEP-1601\\
February 24, 2016
\end{flushright}

\vspace{0.7cm}
\begin{center}
\Large\bf\boldmath
Elements of QED-NRQED Effective Field Theory:\\ I. NLO scattering at leading power
\unboldmath
\end{center}

\vspace{0.8cm}
\begin{center}
{\sc Steven P. Dye, Matthew Gonderinger and Gil Paz}\\
\vspace{0.4cm}
{\it 
Department of Physics and Astronomy \\
Wayne State University, Detroit, Michigan 48201, USA 
}

\end{center}
\vspace{1.0cm}
\begin{abstract}
  \vspace{0.2cm}
  \noindent
The proton radius puzzle, i.e. the large discrepancy in the extraction of the proton charge radius between regular and muonic hydrogen, challenges our understanding of the structure of the proton. It can also be an indication of a new force that couples to muons, but not to electrons. An effective field theory analysis using Non Relativistic Quantum Electrodynamics (NRQED) indicates that the muonic hydrogen result can be interpreted as a large, compared to some model estimates, muon-proton spin-independent contact interaction. The muonic hydrogen result can be tested by a muon-proton scattering experiment, MUSE, that is planned at the Paul Scherrer Institute in Switzerland. The typical momenta of the muons in this experiment are of the order of the muon mass. In this energy regime the muons are relativistic but the protons are still non-relativistic. The interaction between the muons and protons can be described by a hybrid QED-NRQED effective field theory. We present some elements of this effective field theory. In particular we consider ${\cal O}(Z\alpha)$ scattering up to power $m^2/M^2$, where $m$ ($M$) is the muon (proton) mass and $Z=1$ for a proton, and ${\cal O}(Z^2\alpha^2)$ scattering at leading power. We show how the former reproduces Rosenbluth scattering up to power $m^2/M^2$ and the latter the relativistic scattering off a static potential.  Proton structure corrections at ${\cal O}(Z^2\alpha^2)$ will be considered in a subsequent paper. 
\end{abstract}
\vfil

\end{titlepage}

\section{Introduction}
In 2010 the first measurement of the proton charge radius from spectroscopy of muonic hydrogen was found to be five standard deviations away from the value extracted from regular hydrogen \cite{Pohl:2010zza}. More than five years later, this ``proton radius puzzle" is still unresolved, see \cite{Carlson:2015jba} for a recent review. 

The most exciting interpretation of the puzzle is that of a new force that couples to muons and not electrons. But before considering such an option, one would like to rule out a standard model interpretation. Since the proton charge radius can also be extracted from electron-proton scattering, some of the discussion in the literature has focused on reevaluation of the extraction of proton radii from scattering, see for example the $z$-expansion based studies \cite{Hill:2010yb, Epstein:2014zua,Lee:2015jqa}, and references therein\footnote{Some other $z$-expansion based studies do not bound the coefficients of the $z$ expansion \cite{Lorenz:2014vha, Lorenz:2014yda, Griffioen:2015hta} or modify it \cite{Horbatsch:2015qda}. These may result in values that are lower than \cite{Hill:2010yb,Lee:2015jqa}. See \cite{Hill:2010yb} for a discussion of the bounding of the coefficients.}. While leading to a more robust error estimate, the value for the proton charge radius of  \cite{Hill:2010yb,Lee:2015jqa} generally disfavors the muonic hydrogen result. It should be noted that other studies not based on the $z$ expansion listed, e.g., in \cite{Agashe:2014kda}, find values that are consistent with the muonic hydrogen result.

Another possibility is that in the extraction of the charge radius from muonic hydrogen the hadronic uncertainty, i.e. a matrix element or elements that cannot be directly related to experiment,  is underestimated.  The proton charge radius is defined via a ``one-photon" probe of the proton structure. At the level of precision needed to extract it from the muonic hydrogen spectroscopy, ``two-photon" effects must be considered. Estimating these effects is challenging \cite{Hill:2011wy}, even using experimental data. The problem is that only the imaginary part of the two-photon amplitude can be related to experimental data: form factors and structure functions. In order to reconstruct the full amplitude from its imaginary part, one needs a subtracted dispersion relation, which requires the knowledge of a subtraction function. Although there are estimates of this function in the literature, see e.g.  \cite{Nevado:2007dd, Birse:2012eb, Gorchtein:2013yga, Alarcon:2013cba, Tomalak:2014dja}, it cannot be extracted fully from data, which introduces a hadronic uncertainty. It should be noted that other studies \cite{Nevado:2007dd, Alarcon:2013cba} do not use dispersion relation analyses but rather effective field theory techniques.

One way of presenting the problem is by the use of an effective field theory. The typical momentum of the muon in muonic hydrogen is of order $m\alpha\sim 1$ MeV, where $m$ is the muon mass. As a result both the muon and the proton can be described as non-relativistic fields. The appropriate effective field theory is called Non Relativistic Quantum Electrodynamics (NRQED) \cite{Caswell:1985ui}, used successfully for problems in precision QED \cite{Kinoshita:1995mt}. It can also be used to describe proton structure effects in systems like muonic hydrogen. For example, it was first used for that purpose in \cite{Pineda:2002as, Pineda:2004mx} as part of a chain of effective field theories where Heavy Baryon Effective Theory (HBET) is matched onto NRQED and NRQED in turn is matched onto potential NRQED (pNRQED).  If one matches directly onto NRQED as was done in \cite{Hill:2011wy}, proton structure effects in muonic hydrogen at the current level of experimental precision depend on two non-perturbative parameters\footnote{Protons also contribute to hadronic vacuum polarization effects. Hadronic vacuum polarization effects can be incorporated in NRQED by matching onto new photon interaction terms. See for example \cite{Kinoshita:1995mt} for such a matching for perturbative vacuum polarization.}: $c_D$, related to the proton charge radius, and $d_2$,  the coefficient  the spin-independent four-fermion contact interaction between protons and muons. Assuming that the proton charge radius is known from another source, the muonic hydrogen result can be interpreted at face value as a measurement of the contact interaction $d_2$. The value obtained is large compared to some model estimates. 

The muonic hydrogen result can be tested by a muon-proton scattering experiment. Such an experiment, MUSE, is planned at the Paul Scherrer Institute in Switzerland \cite{MUSE}. The typical momentum of the muons in the experiment is of the order of the muon mass, $m\sim100$ MeV. At these energies the muon is relativistic but the proton can still be considered as non-relativistic. The appropriate effective field theory for such kinematics was suggested in \cite{Hill:2012rh}. We refer to it as QED-NRQED effective field theory\footnote{We do not include the pion as a dynamical degree of freedom. The effects of the strong interaction are encoded in the non-perturbative QED-NRQED Wilson coefficients $c_i$ and $b_i$ defined below. Backgrounds from pions are discussed in MUSE technical design report, see \cite{MUSE}.}.  

In order to establish this new effective field theory, we show in this paper how it reproduces some known results. In particular, we consider ${\cal O}(Z\alpha)$ scattering up to power $m^2/M^2$, where $m$ ($M$) is the muon (proton) mass and $Z=1$ for a proton, and ${\cal O}(Z^2\alpha^2)$ scattering at leading power\footnote{We use the factors of $Z$ to keep track of the number of proton-photon interactions.}. We show how the former reproduces Rosenbluth scattering \cite{Rosenbluth:1950yq} and the latter the scattering of a relativistic fermion off a static potential \cite{Dalitz:1951ah, Itzykson:1980rh}. 

While this effective field theory has different contact interactions than those of ``pure" NRQED, one would expect that they can be related to the NRQED contact interactions probed by muonic hydrogen spectroscopy.  The determination of the coefficients of the QED-NRQED contact interactions will be considered in a subsequent paper \cite{inprep}.

The ultimate goal of this program is to calculate the muon-proton cross section in QED-NRQED in terms of quantities such as the proton charge radius and the muon-proton contact interactions. This will allow  to connect muonic hydrogen spectroscopy to muon-proton scattering in a model-independent way. This paper is the first step in the program. 

The paper is structured as follows. In section \ref{sec:Lagrangian} we briefly review the QED-NRQED Lagrangian. In section \ref{sec:alpha1} we present the ${\cal O}(Z\alpha)$ scattering up to power $m^2/M^2$.  In section \ref{sec:alpha2} we present ${\cal O}(Z^2\alpha^2)$ scattering at leading power. We present our conclusions in section \ref{conclusions}. Technical details about the  kinematics and the QED-NRQED amplitude are collected in the appendix.

\section{The Lagrangian}\label{sec:Lagrangian} 
\subsection{NRQED Lagrangian}
The NRQED Lagrangian describes the interaction of non-relativistic, possibly composite, spin-half particle $\psi $ with the electromagnetic field.  Up to and including $1/M^2$,  where $M$ is the mass of the spin-half particle, the NRQED Lagrangian is \cite{Caswell:1985ui, Kinoshita:1995mt}
\begin{equation}\label{Lagrangian2}
{\cal L} = \psi^\dagger\left\{ i D_t +c_2\dfrac{\bm {D}^2}{2M}+c_Fe\dfrac{\bm {\sigma\cdot B}}{2M}+c_De\dfrac{[\bm{\nabla\cdot E}]}{8M^2}+ic_Se\dfrac{\bm{\sigma}\cdot\left(\bm{D\times E}-\bm{E\times D}\right)}{8M^2}\right\}\psi +\cdots ,
\end{equation}
where $D_t=\partial/\partial t+ieA^0$,  $\bm D=\bm\nabla-ie\bm A$, $\bm\sigma$ are the Pauli matrices, and $e$  is the electromagnetic coupling constant\footnote{We follow the conventions of \cite{Kinoshita:1995mt}, although in that paper the NRQED Lagrangian describes an electron. In other words, in this paper we take $e$ to be positive.}. These are the components of $D_\mu=\partial_\mu+ieA_\mu$. The notation $[\bm{\nabla\cdot E}]$ denotes that the derivative is acting only on $\bm E$ and not on $\psi$. For a review see \cite{Paz:2015uga}. The (hidden) Lorentz invariance of the Lagrangian implies that $c_2=1$ \cite{Manohar:1997qy, Heinonen:2012km,Hill:2012rh}. The other Wilson coefficients can be related to the proton electromagnetic form factors
\begin{equation}\label{FF}
\langle p(p')|J_\mu^{\rm em}|p(p)\rangle=\bar u(p')
\left[\gamma_\mu F_1(q^2)+\frac{i\sigma_{\mu\nu}}{2M}F_2(q^2)q^\nu\right]u(p)\,,
\end{equation}
via $c_F=F_1(0)+F_2(0)$,  $c_D=F_1(0)+2F_2(0)+8M^2F_1'(0)$, where $F_1^\prime=dF_1(q^2)/dq^2$, and $c_S=2c_F-F_1(0)$. The latter can also be determined by the hidden Lorentz invariance of the Lagrangian \cite{Manohar:1997qy, Heinonen:2012km,Hill:2012rh}. The NRQED Feynman rules can be extracted from figure 3 of \cite{Kinoshita:1995mt} by multiplying the vertices by $-i$ and the propagators by $i$.

At $1/M^2$ there are operators that couple four spin-half fields\footnote{We use the convention of \cite{Hill:2012rh}, where the operators are suppressed by $1/M^2$ instead of $1/M_\chi M$ of  \cite{Kinoshita:1995mt}. The two are related by a factor of $M_\chi/M$, where $M_\chi$ is the mass of the $\chi$ field.}
\begin{equation}\label{Lagrangian4f}
{\cal L}_{\psi\chi}=\dfrac{d_1}{M^2}\psi^\dagger\sigma^i\psi\chi^\dagger\sigma^i\chi+\dfrac{d_2}{M^2}\psi^\dagger\psi\chi^\dagger\chi+\cdots .
\end{equation}
Here $\chi$ is another NRQED field which can be different from $\psi$. The coefficients $d_1$ and $d_2$ start  at order $\alpha^2$, see \cite{Pineda:2002as, Pineda:2004mx,Hill:2011wy}. The $1/M^2$ NRQED Lagrangian of (\ref{Lagrangian2}) and (\ref{Lagrangian4f}) is enough to describe the proton structure effects relevant to the current precision of muonic hydrogen spectroscopy  \cite{Pineda:2002as, Pineda:2004mx,Hill:2011wy}. In particular, $\chi$ is taken to be an NRQED field for the lepton.  In the following calculations we will only need (\ref{Lagrangian2}) to describe the proton's interactions.  

\subsection{QED-NRQED Lagrangian}

As described in the introduction, we are interested in an effective field theory where the muon is relativistic, while the proton is still non-relativistic. In the following we use the muon mass $m$ as a parameter. Since the discussion also applies to an electron, from now on we refer to $m$ as the lepton mass. 

For the application of QED-NRQED considered in this paper, the interactions of the lepton are described using the usual QED Lagrangian
\begin{equation}\label{LagrangianDirac}
{\cal L}=\bar\ell\, \gamma^\mu\,i\left(\partial_\mu+ieQ_\ell A_\mu \right)\ell-m\bar\ell\ell,
\end{equation}
where $Q_\ell=-1$ for a muon or an electron. We have not included $1/M^2$ operators of  \cite{Hill:2012rh}, since they have Wilson coefficients that start at ${\cal O}(\alpha)$. As a result they only lead to $m^2/M^2$ effects beyond  ${\cal O} (Z\alpha)$ which are not considered in this paper.

The NRQED interaction distinguishes between the time-like ($A^0$) and space-like ($A^i$) components of $A^\mu$. Therefore in a photon exchange between a QED field and an NRQED field the photon polarization will be determined by the NRQED vertex. It is often convenient  to use Coulomb gauge, where the photon propagator is different for time-like and space-like components.  It can be found in, e.g., \cite{Kinoshita:1995mt}.

At 1/$M^2$ we can also have contact interactions  of the form $\psi^\dagger\Sigma\psi\,\bar\ell\Gamma\ell$, where $\Gamma$ is a $4\times 4$ matrix and $\Sigma=1_{2\times 2}, \sigma^i$.  The contact interactions must be even under parity and time reversal.  Since both the unit matrix and the Pauli matrices are even under parity,  $\bar\ell\Gamma\ell$ must be parity even too. This implies eight possible options for $\Gamma$, namely,  $1_{4\times 4},\gamma^0,\sigma^{ij}, \gamma^i\gamma^5$ \cite{Peskin:1995ev}. Since $1_{2\times 2}$ ($\sigma^i$) are even (odd) under time reversal,  $1_{4\times 4}$ and $\gamma^0$ can only be combined with $1_{2\times 2}$, while $\sigma^{ij}$ and $\gamma^i\gamma^5$ can only be combined with $\sigma^i$. For the former we use $\epsilon^{ijk}$ and for the latter $\delta^{ij}$.

An operator of the form $\bar \ell\Gamma\ell$ couples the left-handed and right-handed components of the relativistic lepton field if $\Gamma$ contains an even number of gamma matrices.  As a result one would expect that the Wilson coefficient of such an operator would be proportional to $m$. In other words, we have chiral symmetry in the $m\to 0$ limit. This implies that operators with an even number of gamma matrices should be multiplied by $m/M^3$. At $1/M^2$ we therefore have only two possible contact interactions,  
\begin{equation}\label{contactQN} 
{\cal L}_{\ell\psi}=\dfrac{b_1}{M^2}\psi^\dagger\psi\,\bar \ell\gamma^0\ell+\dfrac{b_2}{M^2}\psi^\dagger\sigma^i\psi\,\bar \ell\gamma^i\gamma^5\ell +{\cal O}\left(1/M^3\right),
\end{equation}
where our notation follows that of \cite{Hill:2012rh}.

In the following we will consider QED-NRQED scattering at ${\cal O}(Z\alpha)$, i.e. one-photon exchange, up to power $m^2/M^2$ and QED-NRQED scattering at ${\cal O}(Z^2\alpha^2)$, i.e. two-photon exchange, at leading power.  As we will see, for both we will only need equations (\ref{Lagrangian2}) and (\ref{LagrangianDirac}). As a result,  $b_1$ and $b_2$ start at  ${\cal O}(Z^2\alpha^2)$. These will be considered in a subsequent paper \cite{inprep}.

\section{QED-NRQED scattering at ${\cal O}(Z\alpha)$}\label{sec:alpha1} 

Our first application is the calculation of the QED-NRQED lepton-proton elastic scattering  $\ell(k)+p(p)\to\ell(k^\prime)+p(p^\prime)$ at ${\cal O}(Z\alpha)$ (for the amplitude) and at power $m^2/M^2$. We will see that the result agrees with the result of the Rosenbluth formula \cite{Rosenbluth:1950yq} up to power $m^2/M^2$.

\begin{figure}
\begin{center}
\includegraphics[scale=1,clip=true,trim=5cm 21cm 13cm 4cm]{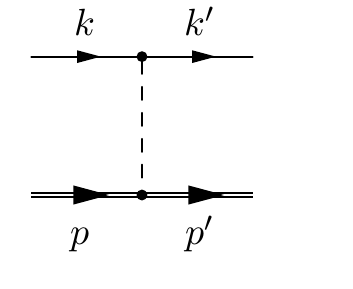}\qquad
\includegraphics[scale=1,clip=true,trim=5cm 21cm 13cm 4cm]{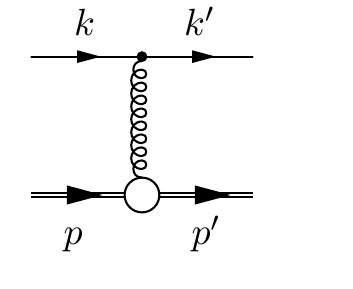}\qquad
\includegraphics[scale=1,clip=true,trim=5cm 21cm 13cm 4cm]{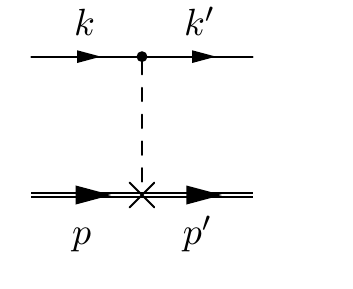}
\caption{\label{figure1} QED-NRQED Feynman diagrams that give a non-zero contribution to elastic lepton-proton scattering at ${\cal O}(Z\alpha)$ up to power $m^2/M^2$. The double line denotes the NRQED field. The dashed (curly) line represents Coulomb (transverse) photon. The dot, circle, and cross vertices represent the Coulomb,  Fermi, and Darwin terms, respectively, see \cite{Kinoshita:1995mt}  for details. 
}
\end{center}
\end{figure}

Calculating the Feynman diagrams of figure \ref{figure1} for a one-photon exchange between a relativistic lepton and a non-relativistic proton up to  $1/M^2$ using (\ref{Lagrangian2}) and (\ref{LagrangianDirac}) we find
\begin{equation}\label{MQN1}
{\cal M}_{\rm QN}=-e^2ZQ_\ell\left[\left(1-c_D\dfrac{\vec q^{\,\,2}}{8M^2}\right)\dfrac1{\vec q^{\,\,2}}\xi_{p^\prime}^\dagger \xi_p\bar u(k^\prime)\gamma^0u(k)+i\dfrac{c_F}{2M}\dfrac1{q^2}\epsilon^{ijk}q^j\xi_{p^\prime}^\dagger \sigma^k\xi_p u(k^\prime)\gamma^iu(k)\right],
\end{equation}
where ``QN"  stands for QED-NRQED, and $\xi_{p^\prime}$ and  $\xi_p$ are two-component spinors. There is no contribution from the operator $\bm {D}^2$ at this order. We have also omitted a contribution from $c_S$ that is proportional to $q_0$ and leads to $1/M^3$ suppressed terms.  

The spin-averaged square of the amplitude can be calculated by an analogue of the Casimir trick, see the appendix.  We find 
\begin{eqnarray}\label{1photonQN} 
\overline{|{\cal M}|}^2_{\rm QN}&=&e^4Z^2Q_\ell^2\left[\dfrac1{\vec q^{\,\,4}}\left(1-c_D\dfrac{\vec q^{\,\,2}}{8M^2}\right)^2\left(4EE^\prime+q^2\right)+\dfrac{c_F^2}{M^2}\dfrac1{q^4}\vec q^{\,\,2}\left(EE^\prime-m^2-\dfrac{\vec k^{\phantom{\prime}}\cdot \vec q\,\,\vec k^\prime\cdot \vec q}{\vec q^{\,\,2}}\right)\right]\nonumber\\
&=&\dfrac{e^4Z^2Q_\ell^2}{\vec q^{\,\,2}}\left[\dfrac1{\vec q^{\,\,2}}\left(4E^2-{\vec q^{\,\,2}}\right)-\dfrac{2E}{M}+\dfrac{\vec q^{\,\,2}+c_F^2\left(\vec q^{\,\,2}+4E^2-4m^2\right)+c_D\left(\vec q^{\,\,2}-4E^2\right)}{4M^2}\right],
\end{eqnarray}
where $E$ ($E^\prime$) is the energy of the initial (final) lepton.  In the second line we have expanded the kinematical variables in powers of $1/M$ and retained only terms up to $1/M^2$, for details see the appendix.  

We can compare this result to Rosenbluth scattering, i.e. the one-photon interaction between a proton, described by the form-factors, and a lepton. Without a considerable increase in complexity,  we can introduce form-factors for the lepton too, since some of the radiative corrections modify the lepton form-factors from the tree-level value of $F_1=1, F_2=0$. We thus have for the lepton-photon vertex  
\begin{equation}
\langle \ell(k')|J_\mu^{\rm em}|\ell(k)\rangle=\bar u(k')
\left[\gamma_\mu F^\ell_1(q^2)-\frac{i\sigma_{\mu\nu}}{2m}F^\ell_2(q^2)q^\nu\right]u(k)\,.
\end{equation} 
The spin averaged square of the amplitude  is given by 
\begin{eqnarray}
\overline{|{\cal M}|}^2&=&\frac{4\pi^2\alpha^2Z^2Q_\ell^2}{q^4}{\rm Tr}\left\{\left(\pslash^\prime+M\right)\left(\gamma_\mu F^p_1(q^2)+\frac{i\sigma_{\mu\alpha}}{2M}F^p_2(q^2)q^\alpha\right)\left(\pslash+M\right)\left(\gamma_\nu F^p_1(q^2)-\frac{i\sigma_{\nu\beta}}{2M}F^p_2(q^2)q^\beta\right)\right\}\nonumber\\
&\times&{\rm Tr}\left\{\left(\kslash^\prime+m\right)\left(\gamma^\mu F^\ell_1(q^2)-\frac{i\sigma^{\mu\rho}}{2m}F^\ell_2(q^2)q_\rho\right)\left(\kslash+m\right)\left(\gamma^\nu F^\ell_1(q^2)+\frac{i\sigma^{\nu\lambda}}{2m}F^\ell_2(q^2)q_\lambda\right)\right\}.
\end{eqnarray}
Collecting the terms by  their powers of $q^2$ we have,  
\begin{eqnarray}\label{1photonFF}
&&\dfrac{\overline{|{\cal M}|}^2}{\pi^2\alpha^2}=\dfrac{256 E^2 (F_1^\ell)^2 (F_1^p)^2 M^2 }{q^4}+\frac{64}{q^2}\bigg[ (F_1^\ell)^2 (F_1^p + F_2^p)^2 m^2 + 
   (F_1^p)^2 (F_1^\ell + F_2^\ell)^2 M^2+\nonumber\\
&+&2 (F_1^\ell)^2 (F_1^p)^2 ME-\frac{E^2\left(\left(F_1^\ell\right)^2 \left(F_2^p\right)^2 m^2 + \left(F_1^p\right)^2 \left(F_2^\ell\right)^2 M^2\right) }{m^2}\bigg]\nonumber\\
&+&16\bigg[ \left(\left(F_1^p\right)^2 + 4 F_1^p F_2^p + \left(F_2^p\right)^2\right) \left(\left(F_1^\ell\right)^2 + 4 F_1^\ell F_2^\ell + \left(F_2^\ell\right)^2\right) +F_1^\ell F_1^p (F_1^\ell F_1^p - 4 F_2^\ell F_2^p)\nonumber\\
&-& \frac{2E \left(\left(F_1^\ell\right)^2 \left(F_2^p\right)^2 m^2 + \left(F_1^p\right)^2 \left(F_2^\ell\right)^2 M^2\right)}{m^2 M}+\frac{E^2 \left(F_2^\ell\right)^2 \left(F_2^p\right)^2}{m^2}\bigg]\nonumber\\
   &+&4q^2\bigg[\frac{F_2^\ell F_2^p \left((2 F_1^\ell + F_2^\ell) F_2^p m^2 + 
    (2 F_1^p + F_2^p) F_2^\ell M^2)\right)}{m^2 M^2}+\frac{2 E \left(F_2^\ell\right)^2 \left(F_2^p\right)^2 }{m^2 M}\bigg]\nonumber\\
&+&q^4\bigg[\frac{\left(F_2^\ell\right)^2 \left(F_2^p\right)^2} {m^2 M^2}\bigg],
\end{eqnarray}
where we have suppressed the dependence of the form factors on $q^2$. Inserting this expression into (\ref{dsigma}) and taking the limit $F_1^\ell\to 1,\, F_2^\ell\to0$ reproduces similar expressions in the literature \cite{Borie:2012tu,Preedom:1987mx}. 
 
As explained in the appendix, in the rest frame of the initial proton, $\overline{|{\cal M}|}^2=4ME_{p^\prime}\overline{|{\cal M}|}^2_{\rm QN}$. Multiplying (\ref{1photonQN}) by $4ME_{p^\prime}$, using the relations $c_F=F_1(0)+F_2(0)$,  $c_D=F_1(0)+2F_2(0)+8M^2F_1'(0)$, and expanding in powers of $1/M$,  we find that the result agrees with the expansion of (\ref{1photonFF}) in powers of $1/M$ in the $F_1^\ell\to 1,\, F_2^\ell\to0$ limit. In particular there is no contribution to the Wilson coefficients of the contact interactions, $b_1$ and $b_2$ at this order. Such contribution arises at  ${\cal O} (Z^2\alpha^2)$ and at power $1/M^2$ and will be considered in a subsequent paper \cite{inprep}.

\section{QED-NRQED scattering at ${\cal O}(Z^2\alpha^2)$ at leading power}\label{sec:alpha2}
We consider elastic lepton-proton scattering $\ell(k)+p(p)\to\ell(k^\prime)+p(p^\prime)$ at ${\cal O}(Z^2\alpha^2)$ at leading power in $m/M$. We will show that the three methods: QED-NRQED at leading power, QED for a point particle at leading power in $1/M$, and scattering off a static $1/r$ potential, give the same  amplitude.    
\subsection{QED-NRQED amplitude}\label{sec:QN2} 
\begin{figure}
\begin{center}
\includegraphics[scale=1,clip=true,trim=5cm 19cm 11cm 4cm]{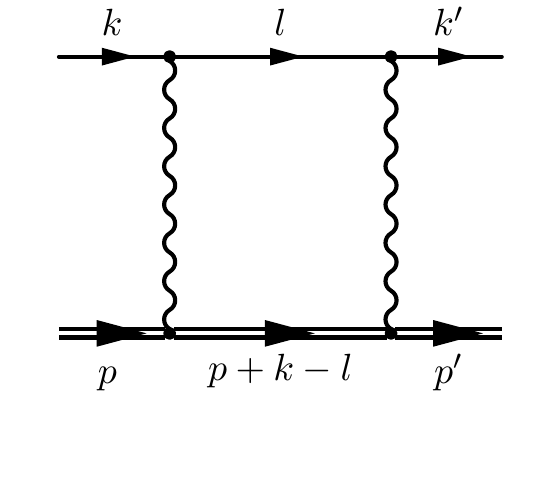}\qquad\qquad
\includegraphics[scale=1,clip=true,trim=5cm 19cm 10cm 4cm]{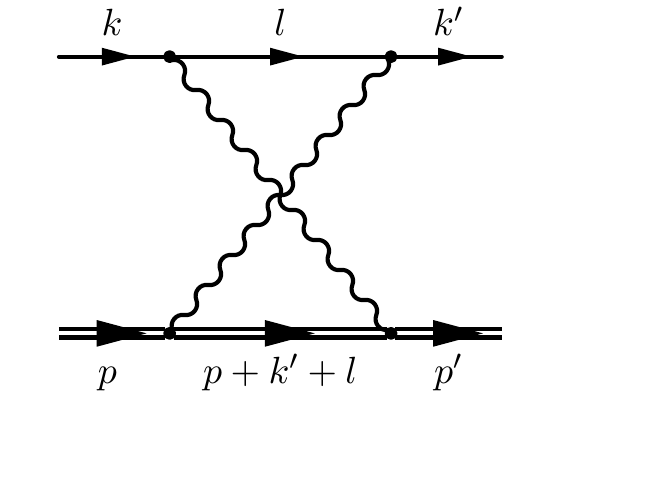}
\caption{\label{figure2} QED-NRQED Feynman diagrams contributing to  elastic lepton-proton scattering at ${\cal O}(Z^2\alpha^2)$ at leading power in $m/M$. The double line denotes the NRQED field.   
}
\end{center}
\end{figure}

The relevant diagrams are shown in figure \ref{figure2}.
The NRQED propagator is $i(p_0-\vec p^{\,2}/2M+i\epsilon)^{-1}$ \cite{Kinoshita:1995mt}. At leading power in $1/M$ we can approximate\footnote{Note that in this approximation the propagator looks like a HQET propagator,  $i(v\cdot p+i\epsilon)^{-1}$, with $v=(1,\vec 0)$. The relation between the HQET and NRQED Lagrangians is discussed in \cite{Manohar:1997qy}.} it as $i(p_0+i\epsilon)^{-1}$. Also, at leading power the NRQED field only couples to $A^0$. Finally, in the rest frame of the proton, $p^0=0$ and $\vec p=0$. The resulting amplitude is (in Feynman gauge) 
\begin{equation}\label{QN2gamma}
{i\cal M}=Z^2Q_\ell^2e^4\int \frac{d^4l}{(2\pi)^4}\dfrac{\bar u(k^\prime)\gamma^0\left(\lslash +m\right)\gamma^0u(k)\xi^\dagger_{p^\prime}\xi_p}{(l-k)^2(l-k^\prime)^2(l^2-m^2)}\left(\dfrac{1}{k^0-l^0+i\epsilon}+\dfrac{1}{l^0-k^{\prime\,0}+i\epsilon}\right). 
\end{equation} 
At the leading power in $1/M$ conservation of momentum and energy imply
\begin{equation}
\sqrt{\vec{k}^2+m^2}+M=\sqrt{\vec{k}^{\prime\,2}+m^2}+\sqrt{M^2+\left(\vec{k}^\prime-\vec k\right)^2}\Rightarrow \sqrt{\vec{k}^2+m^2}=\sqrt{\vec{k}^{\prime\,2}+m^2}+{\cal O}(1/M),
\end{equation}
i.e. $|\vec k|=|\vec{k}^\prime|$ and $k^0=k^{\prime\,0}$. This also implies that $\delta^4(k^\prime+p^\prime-k-p)\approx\delta(k^{\prime\,0}-k^0)\delta^3(\vec k^{\,\prime}+\vec p^{\,\prime}-\vec k)$.  

Using the identity \cite{Gel'fand} $1/\left(x+i\epsilon\right)=P(1/x)-i\pi\delta(x)$, where $P$ is Cauchy principle value, we have at leading power in $1/M$
\begin{equation}
\dfrac{1}{k^0-l^0+i\epsilon}+\dfrac{1}{l^0-k^{\prime\,0}+i\epsilon}=\dfrac{1}{k^0-l^0+i\epsilon}+\dfrac{1}{l^0-k^{0}+i\epsilon}=-2\pi i\delta(l^0-k^0).
\end{equation} 
Averaging over the initial proton spins and summing over the final proton spins implies $\xi^\dagger_{p^\prime}\xi_p\to 1$. Since  $\delta(l^0-k^0)\delta(k^{\prime\,0}-k^0)=\delta(l^0-k^0)\delta(l^0-k^{\prime\,0})$, we can finally write
\begin{eqnarray}\label{QN_Final}
{i\cal M}\,(2\pi)^4\delta^4(k^\prime+p^\prime-k-p)&=&\int \frac{d^4l}{(2\pi)^4}\dfrac{2\pi\delta(l^0-k^0)}{(l-k)^2-\lambda^2}\,\dfrac{2\pi\delta(l^0-k^{\prime\,0})}{(l-k^\prime)^2-\lambda^2}\, \dfrac{\bar u(k^\prime)\gamma^0\left(\lslash +m\right)\gamma^0u(k)}{l^2-m^2}\nonumber\\
&\times&(-)iZ^2Q_\ell^2e^4(2\pi)^3\delta^3(\vec k^{\,\prime}+\vec p^{\,\prime}-\vec k),
\end{eqnarray}
where we have introduced an IR regulator $\lambda$ as the photon ``mass".
\subsection{Point  particle QED amplitude} 
If the proton were a point particle, we could calculate the same diagrams using QED. As we will show, this toy model actually gives the same answer as the effective field theory calculation. The reason is that in the infinite proton mass limit the only information the lepton has about the composite proton is its overall charge. Of course, once we include other properties of the proton such as its magnetic moment or charge radius, described in NRQED by operators suppressed by $1/M$ and $1/M^2$ respectively, the two calculations will differ.        

Calculating the diagrams for a point particle particle of mass $M$ and charge $Ze$ we find
\begin{eqnarray}
{i\cal M}=Z^2Q_\ell^2e^4&\displaystyle{\int}& \frac{d^4l}{(2\pi)^4}\frac1{(l-k)^2-\lambda^2}\frac1{(l-k^\prime)^2-\lambda^2}\dfrac{\bar u(k^\prime)\gamma_\mu\left(\lslash +m\right)\gamma_\nu u(k)}{(l^2-m^2)}\times\nonumber\\
&\times&\bar u(p^\prime)\bigg(\gamma^\mu\dfrac{\pslash+\kslash-\lslash+M}{(p+k-l)^2-M^2}\gamma^\nu +\gamma^\nu\dfrac{\pslash-\kslash^\prime+\lslash+M}{(p-k^\prime+l)^2-M^2}\gamma^\mu \bigg)u(p). 
\end{eqnarray} 
Since $p=(M,\vec 0)$, in the infinite mass limit
 \begin{equation}
 \dfrac{\pslash+\kslash-\lslash+M}{(p+k-l)^2-M^2}\to \dfrac{1+\gamma^0}{2}\cdot\dfrac1{k^0-l^0},\dfrac{\pslash-\kslash^\prime+\lslash+M}{(p-k^\prime+l)^2-M^2}\to \dfrac{1+\gamma^0}{2}\cdot\frac1{l^0-k^{\prime\,0}},
 \end{equation}
 and $u(p)=(\xi_{p},0), \bar u(p^\prime)=(\xi_{p^\prime},0)^\dagger\gamma^0$. As a result $(1-\gamma^0)u(p)=0, \bar u(p^\prime)(1-\gamma^0)=0$. The proton matrix element can be simplified as
 \begin{eqnarray}
 &&\bar u(p^\prime)\gamma^\alpha\left(\dfrac{1+\gamma^0}2\right)\gamma^\beta u(p)= \bar u(p^\prime)\gamma^\alpha\left(\dfrac{1+\gamma^0}2\right)\left(\dfrac{1+\gamma^0}2\right)\gamma^\beta u(p)=\nonumber\\
 &=&\bar u(p^\prime)\left[\left(\dfrac{1-\gamma^0}2\right)\gamma^\alpha+g^{\alpha 0}\right]\left[g^{\beta 0}+\gamma^\beta\left(\dfrac{1-\gamma^0}2\right)\right] u(p)=g^{\alpha 0}g^{\beta 0}\xi_{p^\prime}^\dagger\xi_p. 
 \end{eqnarray}
 All together we obtain 
 \begin{equation}
{i\cal M}=Z^2Q_\ell^2e^4\int \frac{d^4l}{(2\pi)^4}\dfrac{\bar u(k^\prime)\gamma^0\left(\lslash +m\right)\gamma^0u(k)\xi^\dagger_{p^\prime}\xi_p}{(l-k)^2(l-k^\prime)^2(l^2-m^2)}\left(\dfrac{1}{k^0-l^0+i\epsilon}+\dfrac{1}{l^0-k^{\prime\,0}+i\epsilon}\right), 
\end{equation} 
which is the same result as from the QED-NRQED calculation, see equation (\ref{QN2gamma}). We now proceed in the same way as in the previous section to get equation (\ref{QN_Final}).
 
\subsection{Static potential amplitude} 
We consider a lepton scattering off a static external potential \cite{Dalitz:1951ah, Itzykson:1980rh} : 
\begin{equation}
\vec A=0, \quad A^0=\dfrac{Ze\, e^{-\lambda r}}{4\pi r}=-Ze\int \dfrac{d^4q}{(2\pi)^4}\dfrac{2\pi\delta(q^0)}{q^2-\lambda^2}e^{iqx},
\end{equation}
 This implies that in terms of Feynman rules we have a factor of ${2\pi\delta(q^0)}/{(q^2-\lambda^2)}$ for each photon exchange with the potential.  Calculating the transition matrix element we have 
\begin{eqnarray}
{i\cal M}\,(2\pi)\delta(k^{\prime\,0}-k^0)&=&-iZ^2Q_\ell^2e^4\int \frac{d^4l}{(2\pi)^4}\dfrac{2\pi\delta(l^0-k^0)}{(l-k)^2-\lambda^2}\cdot\dfrac{2\pi\delta(l^0-k^{\prime\,0})}{(l-k^\prime)^2-\lambda^2}\cdot \dfrac{\bar u(k^\prime)\gamma^0\left(\lslash +m\right)\gamma^0u(k)}{l^2-m^2}.\nonumber\\
\end{eqnarray}
Up to a factor of $(2\pi)^3\delta^3(\vec k^{\,\prime}+\vec p^{\,\prime}-\vec k)$ this is the same result as the QED-NRQED calculation, equation (\ref{QN_Final}).
\subsection{Cross section}
For completeness we calculate also the cross section. The calculation is similar to \cite{Dalitz:1951ah, Itzykson:1980rh} but the integrals are calculated using the standard Feynman parameters.  We start from equation (\ref{QN_Final}). Using
\begin{equation}
\dfrac{\delta(l^0-k^0)\delta(l^0-k^{\prime\,0})}{\left[(l-k)^2-\lambda^2\right]\left[(l-k^\prime)^2-\lambda^2\right]\left[l^2-m^2\right]}=\dfrac{\delta(l^0-k^0)\delta(l^0-k^{\prime\,0})}{\left[(\vec l-\vec k)^2+\lambda^2\right]\left[(\vec l-\vec k^\prime)^2+\lambda^2\right]\left[\vec k^2-\vec{l}^2\right]}\,,
\end{equation}
and 
\begin{equation}
\delta(l^0-k^0)\bar u(k^\prime)\gamma^0\left(\lslash +m\right)\gamma^0u(k)=\delta(l^0-k^0)\bar u(k^\prime)\left(k^0\gamma^0+m+\vec l\cdot \vec \gamma\right)u(k), 
\end{equation}
we get 
\begin{eqnarray}\label{MQN2}
{i\cal M}\,(2\pi)^4\delta^4(k^\prime+p^\prime-k-p)&=&-2i\dfrac{Z^2Q_\ell^2\alpha^2}{\pi}2\pi\delta(k^0-k^{\prime\,0})(2\pi)^3\delta^3(\vec k^{\,\prime}+\vec p^{\,\prime}-\vec k)\times\nonumber\\
&\times&\int d^3l\,\dfrac{\bar u(k^\prime)\left(k^0\gamma^0+m+\vec l\cdot \vec \gamma\right)u(k)}{\left[(\vec l-\vec k)^2+\lambda^2\right]\left[(\vec l-\vec k^\prime)^2+\lambda^2\right]\left[\vec k^2-\vec{l}^2+i\epsilon\right]}.
\end{eqnarray}
We need two integrals
\begin{eqnarray}\label{integrals}
I_1&=&\int d^3l\,\dfrac{1}{\left[(\vec l-\vec k)^2+\lambda^2\right]\left[(\vec l-\vec k^\prime)^2+\lambda^2\right]\left[\vec k^2-\vec{l}^2+i\epsilon\right]},\nonumber\\
I_2^i&=&\int d^3l\,\dfrac{l^i}{\left[(\vec l-\vec k)^2+\lambda^2\right]\left[(\vec l-\vec k^\prime)^2+\lambda^2\right]\left[\vec k^2-\vec{l}^2+i\epsilon\right]}.
\end{eqnarray}
The denominators arising from the photon propagators can be combined using a Feynman parameter as 
\begin{equation}
x\left[(\vec l-\vec k)^2+\lambda^2\right]+\bar x\left[(\vec l-\vec k^\prime)^2+\lambda^2\right]=(\vec l-\vec K)^2+M^2,
\end{equation}
where $0\leq x\leq 1$, $\bar x=1-x$, $\vec K=x\vec k+\bar x\vec k^\prime$, $M^2=-\vec K^2+\vec k^2+\lambda^2$, and we have used $\vec k^2=\vec k^{\prime\,2}$. Combining this with the third denominator of (\ref{integrals}) using another Feynman parameter we find
\begin{eqnarray}
I_1&=&-2\int_0^1 dx\int_0^1 dy\, y\int d^3 l \dfrac1{\left(\vec l^2+\Delta-i\epsilon\right)^3}  \nonumber\\
I_2^i&=&-2\int_0^1 dx\int_0^1 dy\, y^2\int d^3 l \dfrac{K^i}{\left(\vec l^2+\Delta-i\epsilon\right)^3}\,, 
\end{eqnarray}
where $\Delta=y\bar y\vec K^2+yM^2-\bar y\vec k^2$ and we have changed $\vec l\to \vec l-\vec K y$. It is convenient to perform the integral over $|\vec l|$ first and then to integrate over $y$.  For the $x$ integral we note that $\Delta$ is a function of $x(1-x)$. We split the integration range into two intervals, $0\leq x\leq 1/2$ and  $1/2\leq x\leq 1$, and change variables to $z=x(1-x)$. Thus for a function $f(x)$, 
\begin{equation}
\int_0^1 dx\, f(x)= \int_0^{\frac14} dz\, \dfrac{f\left(\frac12-\frac12\sqrt{1-4z}\right)+f\left(\frac12+\frac12\sqrt{1-4z}\right)}{\sqrt{1-4z}}.
\end{equation}
After the change of variables, $\vec K^2=\vec k^2-4\vec k^2z\sin^2\frac{\theta}2$ and $M^2=\lambda^2+4\vec k^2z\sin^2\frac{\theta}2$. Performing the $|\vec l|$ and $y$ integrations we have
\begin{equation}
I_1=-2\int_0^\frac14 \frac{dz}{\sqrt{1-4z}}\,\dfrac{\pi^2}{M\left[\left(M-i|\vec k|\right)^2+\vec K^2\right]}
\end{equation}
\begin{eqnarray}
I_2^i=-2\pi^2\left(\dfrac{k^i+k^{\prime i}}{2}\right)\int_0^\frac14&&\hspace{-1em}\frac{dz}{\sqrt{1-4z}}\Bigg\{\dfrac1{M\vec K^2}+\dfrac{iM|\vec K|+\vec k^2}{M\vec K^2\left[\left(M-i|\vec k|\right)^2+\vec K^2\right]}+\nonumber\\
&+&\dfrac{i}{2|\vec K|^3}\log\left(\frac{i M +|\vec k|-|\vec K|}{i M +|\vec k|+|\vec K|}\right)\Bigg\}.
\end{eqnarray}
The polynomial terms in $I_1$ and $I_2^i$ can be integrated directly. For the logarithmic term in $I_2^i$ it is convenient to use integration by parts. Defining $I_2^i\equiv I_2\left(k^i+k^{\prime i}\right)/2$, we find
\begin{eqnarray} 
I_1&=&\dfrac{\pi^2}{2i|\vec k|^3 \sin^2\frac{\theta}2} \log\left(\dfrac{2|\vec k|\sin\frac{\theta}2}{\lambda}\right)\nonumber\\
I_2&=&\dfrac{\pi^2}{2|\vec k|^3 \cos^2\frac{\theta}2}\left\{\dfrac{\pi}2\left(1-\dfrac1{\sin\frac{\theta}2}\right)-i\left[\frac1{\sin^2\frac{\theta}2}\log\left(\dfrac{2|\vec k|\sin\frac{\theta}2}{\lambda}\right)+\log\dfrac{\lambda}{2|\vec k|}\right]\right\}.
\end{eqnarray}
This is the same result of \cite{Itzykson:1980rh}. As was pointed out in \cite{Hill:2012rh}, \cite{Dalitz:1951ah} has the wrong sign for $I_1$. 

Since $\kslash u(k)=(k^0\gamma^0-\vec k\cdot \vec\gamma)u(k)=mu(k)$, we have $\vec k\cdot \vec\gamma\, u(k)=(k^0\gamma^0-m)u(k)$. Similarly $\bar u(k^\prime)\,\vec k^\prime\cdot\vec\gamma=\bar u(k^\prime)(k^{\prime\,0}\gamma^0-m)$. Equation (\ref{MQN2}) simplifies to
\begin{equation}
{\cal M}^{(2)}_{\rm QN}=-\dfrac{2Z^2Q_\ell^2\alpha^2}{\pi}\bar u(k^\prime)\left[m(I_1-I_2)+k^0\gamma^0(I_1+I_2)\right]u(k),
\end{equation}
where we have added the subscript ``QN" to denote that we are using non-relativistic normalization for the proton states.  

The ${\cal O} (Z\alpha)$ amplitude at leading power is obtained from equation (\ref{MQN1}) by keeping only the leading power term and replacing  $\xi^\dagger_{p^\prime}\xi_p\to 1$, see section \ref{sec:QN2}. We have
\begin{equation}
{\cal M}^{(1)}_{\rm QN}=-4\pi\alpha ZQ_\ell\dfrac1{\vec q^{\,\,2}}u(k^\prime)\gamma^0u(k).
\end{equation}
At leading power in $1/M$ the relation between ${\cal M}_{\rm QN}$ and ${\cal M}$ in the initial proton rest frame is just ${\cal M}=2M{\cal M}_{\rm QN}$, see the appendix, and we obtain 
\begin{equation}
{\cal M}^{(1+2)}=\dfrac{-8M\pi\alpha ZQ_\ell}{\vec q^{\,\,2}}\bar u({k^\prime})\left\{\gamma^0\left[1+\alpha Z Q_\ell\dfrac{k^0\vec q^{\,\,2}}{2\pi^2}(I_1+I_2)\right]+\alpha Z Q_\ell\dfrac{m\vec q^{\,\,2}}{2\pi^2}(I_1-I_2)\right\}.
\end{equation}
At leading power in $1/M$ the cross section is given by $d\sigma/d\Omega=\overline{|{\cal M}|}^2/(64\pi^2M^2)$. We find \begin{eqnarray}
\dfrac{d\sigma}{d\Omega}=\dfrac{4Z^2\alpha^2Q^2_\ell E^2\left(1-v^2{\sin^2{\textstyle\frac{\theta}2}}\right)}{\vec q^{\,\,4}}\left[1+\alpha Z Q_\ell\frac{\vec q^{\,\,2}E}{\pi^2}\left(\mbox{Re}\,(I_1+I_2)+\dfrac{m^2\,\mbox{Re}\,(I_1-I_2)}{E^2\left(1-v^2{\sin^2\frac{\theta}2}\right)}\right)\right]\!,
\end{eqnarray}
where $E=k^0$ and $v=|\vec k|/k^0$. Since $I_1$ is purely imaginary, only $I_2$ contributes to the cross section. In particular the dependance on $\lambda$ cancels.  The cross section is finally 
\begin{eqnarray}\label{XS}
\dfrac{d\sigma}{d\Omega}=\dfrac{4Z^2\alpha^2Q^2_\ell E^2\left(1-v^2{\sin^2{\textstyle\frac{\theta}2}}\right)}{\vec q^{\,\,4}}\left[1-\alpha Z Q_\ell\frac{\pi v\sin{\textstyle\frac{\theta}2}(1-\sin{\textstyle\frac{\theta}2})}{1-v^2{\sin^2{\textstyle\frac{\theta}2}}}\right].
\end{eqnarray}
Taking $Q_\ell=-1$ we obtain the results\footnote{Note that \cite{Dalitz:1951ah} uses $A^0=Ze\, e^{-\lambda r}/r$. As a result, one needs to replace $\alpha\to e^2$ in the comparison. Also, one has to be careful about the relative sign between the lepton and the potential charges in \cite{Itzykson:1980rh}.} of \cite{Dalitz:1951ah, Itzykson:1980rh}.  

\subsection{Anti-lepton cross section}
In the calculation above we have assumed that the lepton is a particle. It is instructive to see how (\ref{XS}) changes for anti-lepton-proton scattering. The answer, ``Take $Q_\ell=+1$ in (\ref{XS})" is correct, but since for QED the Feynman rule for the vertex is the same for leptons and anti-leptons, it is not immediately obvious why this is true. Beyond the theoretical interest, MUSE will consider both $\mu^\pm p$ and $e^\pm p$ scattering \cite{MUSE}, so it is instructive to see how the cross section changes. 

Ignoring overall minus signs, apart from sign difference between lepton and anti-leptons,  the leptonic part of the ${\cal O}(Z\alpha)$ amplitude is given by 
\begin{eqnarray}
{\cal M}^{(1)}_{\ell^-}&=&Z\alpha\,\bar u(k^\prime)\gamma^\mu u(k)A_\mu(k-k^\prime)\dots\nonumber\\
{\cal M}^{(1)}_{\ell^+}&=&-Z\alpha\,\bar v(k)\gamma^\mu v(k^\prime)A_\mu(k-k^\prime)\dots\,.  
\end{eqnarray} 

\begin{figure}
\begin{center}
\includegraphics[scale=1,clip=true,trim=5cm 22cm 11cm 3cm]{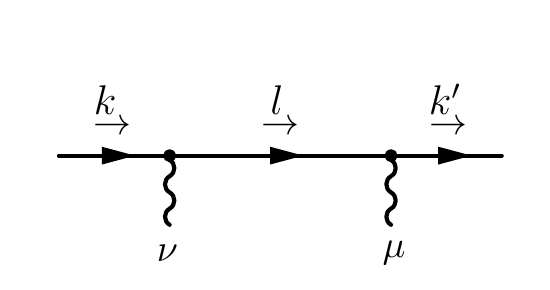}\qquad\qquad
\includegraphics[scale=1,clip=true,trim=5cm 22cm 10cm 3cm]{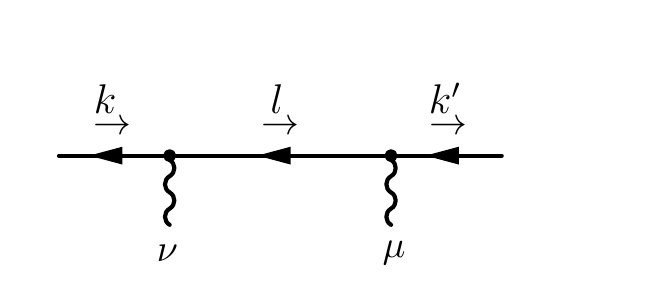}
\caption{\label{figure3} The leptonic part of the ${\cal O}(Z^2\alpha^2)$ amplitude at leading power in $m/M$ for a lepton (left) and an anti-lepton (right). 
}
\end{center}
\end{figure}

As seen in figure \ref{figure3}, the leptonic part of the ${\cal O}(Z^2\alpha^2)$ amplitude is 
\begin{eqnarray}
{\cal M}^{(2)}_{\ell^-}&=&Z^2\alpha^2\int \frac{d^4 l}{(2\pi)^4}\bar u(k^\prime) \gamma^\mu\frac{\left(\lslash+m\right)}{l^2-m^2}\gamma^\nu u(k)A_\mu(l-k^\prime)A_\nu(k-l)\dots\nonumber\\
{\cal M}^{(2)}_{\ell^+}&=&-Z^2\alpha^2\int \frac{d^4 l}{(2\pi)^4}\bar v(k) \gamma^\nu\frac{\left(-\lslash+m\right)}{l^2-m^2}\gamma^\mu v(k^\prime)A_\mu(l-k^\prime)A_\nu(k-l)\dots\,.  
\end{eqnarray} 

Notice that ${\cal M}^{(1)}_{\ell^+}$ and  ${\cal M}^{(2)}_{\ell^+}$ have the same overall sign.   Calculating the spin-averaged leptonic part of the squared amplitude we have the following traces 
\begin{eqnarray}
\mbox{Leptons:}&& \mbox{Tr}\left\{\left(\kslash^\prime+m\right)\left[\gamma^\rho+Z\alpha\gamma^\mu\left(a\,\lslash+b\,m\right)\gamma^\nu\right]\left(\kslash+m\right)\left[\gamma^{\rho^\prime}+Z\alpha\gamma^{\nu^\prime}\left(a^*\,\lslash+b^*\,m\right)\gamma^{\mu^\prime}\right]\right\}\nonumber\\
\mbox{Anti leptons:}&& \mbox{Tr}\left\{\left(\kslash-m\right)\left[\gamma^\rho+Z\alpha\gamma^\nu\left(-a\,\lslash+b\,m\right)\gamma^\mu\right]\left(\kslash^\prime-m\right)\left[\gamma^{\rho^\prime}+Z\alpha\gamma^{\mu^\prime}\left(-a^*\,\lslash+b^*\,m\right)\gamma^{\nu^\prime}\right]\right\}\,,\nonumber\\
\end{eqnarray}
where $a$ and $b$ contain integrals over $d^4l$ and we ignore overall factors common to the two traces. Collecting the terms arising from the inference between ${\cal M}^{(1)}$ and ${\cal M}^{(2)}$, i.e. the ${\cal O}(Z^3\alpha^3)$  terms in the cross section, we always pick up even number of gamma matrices which imply we always get an extra minus sign for the anti-leptons. The order of the gamma matrices also changes, but because of the symmetries of trace, this has no effect. The cross section is therefore, 
\begin{eqnarray}
\dfrac{d\sigma_{\ell^\mp}}{d\Omega}=\dfrac{4Z^2\alpha^2E^2\left(1-v^2{\sin^2{\textstyle\frac{\theta}2}}\right)}{\vec q^{\,\,4}}\left[1\pm\alpha Z \frac{\pi v\sin{\textstyle\frac{\theta}2}(1-\sin{\textstyle\frac{\theta}2})}{1-v^2{\sin^2{\textstyle\frac{\theta}2}}}\right].
\end{eqnarray}

\section{Conclusions and outlook}\label{conclusions}
It has been almost six years since the first measurement of the proton charge radius in muonic hydrogen was published \cite{Pohl:2010zza} and found to be five standard deviations away from the regular hydrogen value. In the intervening time many studies have looked into the extraction of the radius from regular and muonic hydrogen spectroscopy as well as from scattering, see \cite{Carlson:2015jba} for a recent review, but this ``proton radius puzzle" is still unresolved. 

One of the issues involved in the extraction of the proton charge radius from muonic hydrogen is  the hadronic uncertainty associated with the two-photon exchange amplitude. Only its imaginary part can be directly reconstructed from experimental data. Due to the need for subtraction in the dispersion relation, the amplitude cannot be fully reconstructed from its imaginary part. We have some information about the subtraction function, but  by and large, it has to be modeled \cite{Hill:2011wy}.  

There have been several studies of this issue,  see e.g.  \cite{Nevado:2007dd, Birse:2012eb, Gorchtein:2013yga, Alarcon:2013cba, Tomalak:2014dja}, but considering the far-reaching implications of the puzzle it is important to explore a variety of  approaches. One such approach is to directly match onto NRQED to describe proton structure effects in hydrogen-like systems as was done in \cite{Hill:2011wy}\footnote{See also \cite{Pineda:2002as, Pineda:2004mx} for a different approach that first used NRQED for this problem.}.  From such an analysis one finds that the muonic hydrogen measurement depends on two Wilson coefficients in the NRQED Lagrangian. One is equivalent to the charge radius. The other is the coefficient of the spin-independent muon-proton contact  interaction and could be determined by matching to the two-photon amplitude, if it was fully known.  

The muonic hydrogen result  can be tested in the planned muon-proton scattering experiment, MUSE \cite{MUSE}. In this experiment the energy of the muons is of the order of the muon mass. As a result, the appropriate effective field theory has relativistic muons but non-relativistic protons. Such an effective field theory, QED-NRQED, was suggested in \cite{Hill:2012rh}. Lepton-proton scattering in QED-NRQED is naturally organized as an expansion in $\alpha$ and $m/M$. In this paper we presented two QED-NRQED calculations: ${\cal O}(Z\alpha)$ corrections to the amplitude up to and including power $m^2/M^2$ and ${\cal O}(Z^2\alpha^2)$ at leading power in $m/M$. 

QED-NRQED lepton-proton scattering at ${\cal O}(Z\alpha)$ and power $m^2/M^2$  reproduces the known Rosenbluth scattering formula, i.e. the one-photon exchange cross section expressed in terms of the proton form factors \cite{Rosenbluth:1950yq}, up to power $m^2/M^2$. It requires just the Dirac Lagrangian and the NRQED Lagrangian up to $1/M^2$. In particular there is no contribution at this order from $1/M^2$ corrections to the Dirac Lagrangian \cite{Hill:2012rh} and more importantly from the lepton-proton contact interactions. This implies that the coefficients of these operators start at a higher order in $\alpha$. In particular, one would expect that the first non-zero contribution to $b_1$ and $b_2$ in equation (\ref{contactQN})  would be at ${\cal O}(Z^2\alpha^2)$ . For that, one has to calculate an appropriate amplitude to ${\cal O}(Z^2\alpha^2)$ and power $m^2/M^2$ and will be considered in a subsequent paper \cite{inprep}. 

QED-NRQED lepton-proton scattering at ${\cal O}(Z^2\alpha^2)$ and at leading power reproduces the ${\cal O}(Z^2\alpha^2)$ terms in the scattering of a lepton off a static $1/r$ potential \cite{Dalitz:1951ah, Itzykson:1980rh}. Interestingly it also reproduces the lepton scattering off a ``point particle" proton at leading power in $1/M$. It is easy to understand why. In the $M\to \infty$ limit  the only information the lepton has about the proton is the proton's charge $Ze$. Effects such as the proton magnetic moment and the proton charge radius arise only at $1/M$ and $1/M^2$ respectively, see equation (\ref{Lagrangian2}). QED-NRQED can naturally incorporate such effects.  For completeness we have also calculated the cross section, but unlike \cite{Dalitz:1951ah, Itzykson:1980rh} we used the standard technique of Feynman parameters. Still, these leading power integrals are not representative of the typical integrals one would obtain in calculating QED-NRQED diagrams at higher powers. We will discuss such integrals in a subsequent paper \cite{inprep}. Finally, we have also commented on the change in the cross section when we consider anti-lepton scattering. 

These calculations validate QED-NRQED and set the stage for its use in addressing the proton radius puzzle. The next step will be to relate the Wilson coefficients $b_1$ and $b_2$ to the full two-photon amplitude and to the NRQED Wilson coefficients $d_1$ and $d_2$ either directly by matching, or indirectly via the full two-photon amplitude\footnote{An analogous, but different, matching between HBET and NRQED was done in \cite{Nevado:2007dd}.}.  Once this is done, one could calculate the lepton-proton cross section in QED-NRQED. Ideally this would lead to a direct model-independent relation between muon-proton scattering and muonic hydrogen spectroscopy, or in other words, use data to resolve the hadronic uncertainty.

\vskip 0.2in
\noindent
{\bf Acknowledgements}
\vskip 0.1in
\noindent
We  thank Andrew E. Blechman and Alexey A. Petrov for useful discussions and  comments on the manuscript. We also thank Richard J. Hill for useful discussions. This work was supported by NIST Precision Measurement Grants Program and DOE grant DE-SC0007983.

\begin{appendix}
\section{Appendix}\label{appendix}
\label{app:A}
\subsection{Kinematics}
We consider lepton-proton scattering, $\ell(k)+p(p)\to \ell(k^\prime)+p(p^\prime)$, in the initial proton rest frame, i.e.  $\vec p=0$. We denote the lepton mass by $m$ and the proton mass by $M$. The initial lepton energy is $E$ and the final lepton energy is $E^\prime$. The scattering angle, i.e. the angle between $\vec{k}$ and $\vec{k^\prime}$ is $\theta$.  We define $q=k-k^\prime=p^\prime-p$. 

For spin-averaged $2\to 2$ scattering there are only two independent variables, so many of the kinematical variables  can be related to one another:
\begin{eqnarray}
&& p^\prime=p+q, \quad k^\prime=k-q,\quad  p^2=M^2, \quad k^2=m^2,\nonumber\\
&&p\cdot q=M(E-E^\prime)=Mq^0=-q^2/2,\nonumber\\
&&k\cdot q=q^2/2,\quad {\vec q}^{\,\,2}=-q^2+q^4/4M^2.
\end{eqnarray}

There are also several approximate relations between the various kinematic variables:
\begin{eqnarray}
q^2&=&-{\vec q}^{\,\,2}+\vec q^{\,\,4}/4M^2+{\cal O}\left(\frac1{M^3}\right),\nonumber\\
\vec k\cdot \vec q&=&{\vec q}^{\,\,2}/2+{\cal O}\left(\frac1{M}\right),\nonumber\\
\vec k^\prime\cdot \vec q&=&-{\vec q}^{\,\,2}/2+{\cal O}\left(\frac1{M}\right).
\end{eqnarray}

 The differential cross section is given by:
\begin{equation}\label{dsigma}
\frac{d\sigma}{d \Omega}=\frac{1}{64\pi^2 M} \frac{|\vec k^\prime |}{|\vec k|}\frac{1}{\left|M+E-\dfrac{|\vec k| E^\prime\cos \theta}{|\vec k^\prime |}\right|}\overline{|{\cal M}|}^2,
\end{equation}
where as usual $\overline{|{\cal M}|}^2$ is the spin-averaged amplitude squared. 
\subsection{QED-NRQED amplitude}
Usually the Dirac spinors are normalized via $u^\dagger u=2E$. For NRQED the spinors are normalized as $\xi^\dagger\xi=1$. As a result we can relate the amplitude for lepton-proton scattering in the standard normalization (${\cal M}$) to that of QED-NRQED (${\cal M}_{\rm QN}$) via  ${\cal M}=2\sqrt{E_{p^\prime}E_p}\,{\cal M}_{\rm QN}$. In the rest frame of the initial proton the spin averaged amplitudes $\overline{|{\cal M}|}^2$ and $\overline{|{\cal M}|}^2_{\rm QN}$ are related by $\overline{|{\cal M}|}^2=4ME_{p^\prime}\overline{|{\cal M}|}^2_{\rm QN}$, where $E_{p^\prime}=\sqrt{M^2+\vec q^{\,\,2}}$.

Spin averaged squared amplitudes in QED-NRQED can be calculated by an analogue of the Casimir trick. Thus for the amplitude of the form ${\cal M}=\xi_{p^\prime}^\dagger\,\Sigma\, \xi_p\, \bar u(k^\prime)\, \Gamma\, u(k)$, where $\xi$  is a two-component  spinor, $\Sigma=\vec \sigma$ or $1_{2\times2}$, $u$ is a Dirac spinor, and $\Gamma$ part of the Dirac basis,  
\begin{equation}
\overline{|{\cal M}|}^2_{\rm QN}=\frac14\mbox{Tr}\left[\Sigma\Sigma^\dagger\right]\mbox{Tr}\left[(\kslash+m)\Gamma(\kslash^\prime+m)\overline\Gamma\,\right],
\end{equation}
where $\overline\Gamma=\gamma^0\Gamma^\dagger\gamma^0$.
\end{appendix}

\end{document}